\documentclass[12pt]{article}
\usepackage{color}
\usepackage{amssymb,latexsym,bm, amsmath,amsthm, amsfonts,multirow, graphicx,mathrsfs}
\usepackage{eufrak,booktabs}
\usepackage{epsfig}
\usepackage{enumitem}
\usepackage{verbatim}

\newcommand{\argmin}{\mathop{\rm argmin}}
\newcommand{\argmax}{\mathop{\rm argmax}}

\def\Ab{{\boldsymbol A}}
\def\Bb{{\boldsymbol B}}
\def\Cb{{\boldsymbol C}}

\def\Ib{{\boldsymbol I}}

\def\Kb{{\boldsymbol K}}

\def\Mb{{\boldsymbol M}}

\def\Pb{{\boldsymbol P}}

\def\Vb{{\boldsymbol V}}

\def\Xb{{\boldsymbol X}}

\def\etab{{\boldsymbol        \eta}}

\def\Deltab{{\boldsymbol      \Delta}}

\def\Sigmab{{\boldsymbol      \Sigma}}

\def\vec{{\rm vec}}

\def\1{{\bm 1}}
\def\0{{\bm 0}}

\def\Xb{{\bm X}}

\renewcommand{\hat}{\widehat}

\newtheorem{thm}{Theorem}

\newtheorem{rmk}[thm]{Remark}

\begin{document}

\title{A low-rank based estimation-testing procedure
for matrix-covariate regression}
\author{Hung Hung and Zhi-Yu Jou\\
\small Institute of Epidemiology and Preventive Medicine\\
\small National Taiwan University, Taiwan}
\date{}
\maketitle

\begin{abstract}
Matrix-covariate is now frequently encountered in many biomedical
researches. It is common to fit conventional statistical models by
vectorizing matrix-covariate. This strategy, however, results in a
large number of parameters, while the available sample size is
relatively too small to have reliable analysis results. To overcome
the problem of high-dimensionality in hypothesis testing, variance
component test has been proposed with promise detection power, but
is not straightforward to provide estimates of effect size. In this
work, we overcome the problem of high-dimensionality by utilizing
the inherent structure of the matrix-covariate. The advantage is
that estimation and hypothesis testing can be conducted
simultaneously as in the conventional case, while the estimation
efficiency and detection power can be largely improved, due to a
parsimonious parameterization for the coefficients of
matrix-covariate. Our method is applied to test the significance of
gene-gene interactions in the PSQI data, and is applied to test if
electroencephalography is associated with the alcoholic status in
the EEG data, wherein sparse effects and low-rank effects of
matrix-covariates are identified, respectively.

\end{abstract}

\clearpage

\section{Introduction}\label{sec.1}

Matrix-covariate is now frequently encountered in many biomedical
researches. Let $Y$ be the response of interest and $\Mb\in
\mathbb{R}^{p\times q}$ be the $p\times q$ matrix-covariate. The
research aim focuses on the association between $Y$ and $\Mb$,
possibly after adjusting the effects of confounding factors $Z\in
\mathbb{R}^m$. In the Electroencephalography (EEG) data, for
example, $Y$ is the binary alcoholic status, and $\Mb$ is a
$256\times 64$ matrix with its $(j,k)$-th element $\Mb(j,k)$ being
the voltage value of the $k$-th electrode measured at the $j$-th
time point. Sometimes, $\Mb$ is not directly observed but is induced
from the original covariates. In the PSQI data, for instance, it is
of interest to test if the PSQI score $Y$ is associated with
gene-gene interactions (G$\times$G) among one $p$-genetic markers
$G=(g_1,\ldots g_p)^T$ and another $q$-genetic markers
$E=(e_1,\ldots e_q)^T$, i.e., the $pq$ products $\{g_je_k:1\le j\le
p,1\le k\le q\}$. In this case, it is equivalent to test if $Y$ is
associated with the matrix-covariate $\Mb=GE^T$ after adjusting for
the effects of $Z=(G^T,E^T)^T$, by observing that $\Mb(j,k)=g_je_k$.

Two questions are commonly raised by practitioners:
\begin{enumerate}
\item[(Q1)]
Does $Y$ associate with $\Mb$?

\item[(Q2)]
How does $\Mb$ affect $Y$?
\end{enumerate}
To answer (Q1)-(Q2), a simple method is to fit a GLM for $Y$ on each
element of $\Mb$, and then use certain method to combine these $pq$
analysis results. This marginal method, however, can produce biased
results due to the ignorance of the joint effects of $\Mb$. It can
also give a low detection power due to a severe penalty from
adjusting for multiple hypothesis testing. Joint inference is thus
preferable, which fits the GLM
\begin{eqnarray}
Y|(Z,\Mb)\sim {\rm Normal}(E(Y|Z,\Mb),\sigma^2)\label{glm.normal}
\end{eqnarray}
for continuous $Y$ with a common variance $\sigma^2$, or
\begin{eqnarray}
Y|(Z,\Mb)\sim {\rm Bernoulli}( E(Y|Z,\Mb))\label{glm.binary}
\end{eqnarray}
for binary $Y$, and assumes that
\begin{eqnarray}
g\{E(Y|Z,\Mb)\}=\gamma+\xi^TZ+\sum_{j,k}\eta_{jk}\Mb(j,k),\label{model0}
\end{eqnarray}
where $g$ is the link function, $\gamma$ is the intercept term,
$\xi$ is the effect of $Z$, and $\eta_{jk}$ is the effect of
$\Mb(j,k)$. Following the convention, we adopt the identity link
$g(u)=u$ for model~(\ref{glm.normal}), and adopt the logit link
$g(u)=\ln\frac{u}{1-u}$ for model~(\ref{glm.binary}). Based
on~(\ref{model0}), inference about the association between $Y$ and
$\Mb$ relies on the estimation of $\eta_{jk}$ or testing the null
hypothesis
\begin{eqnarray}
H_0:\eta_{jk}=0~~\forall~~(j,k). \label{h0}
\end{eqnarray}
Although the joint method overcomes the problem of bias, it suffers
the problem of high-dimensionality, since the number of parameters
$1+m+pq$ can be large in comparison with the sample size $n$. As a
result, statistical inference procedure can be unstable and
inefficient, which further decreases the detection power to testing
(\ref{h0}).

%

To overcome the problem of high-dimensionality in testing
(\ref{h0}), Lin \emph{et al}. (2013) apply the variance component
test (Lin, 1997) to propose GESAT. The authors extend model
(\ref{model0}) to the generalized linear mixed model (GLMM) by
assuming that $\eta_{jk}$'s independently follow an arbitrary
distribution with zero mean and a common variance $\tau^2$. As a
result, testing (\ref{h0}) is equivalent to testing $H_0:\tau^2=0$
under the GLMM, and the variance component test is applied. The
authors show by simulations that GESAT has higher detection power
than the conventional minimum p-value test. GESAT has the advantage
of fast computation and is shown to be locally most powerful. GESAT,
however, aims to test if $Y$ is associated with $\Mb$ or not (i.e.,
answer (Q1)), but is not straightforward to provide estimates of the
effect sizes of $\Mb$ (i.e., answer (Q2)). A naive solution is to
fit model~(\ref{model0}) to obtain the estimates of $\eta_{jk}$'s,
but it will again suffer the problem of high-dimensionality, and
there is no guarantee that the estimates coincide with the
conclusion from GESAT, either. The aim of this study is thus to
propose a unified inference procedure to answer (Q1)-(Q2) based on
the joint model~(\ref{model0}), while overcoming the problem of
high-dimensionality.

The rest of this article is organized below. Our improved inference
procedure for $\eta_{jk}$'s is developed in Section~2, based on
which two powerful test statistics for (\ref{h0}) are proposed in
Section~3. Section~4 conducts numerical studies to evaluate the
performances of our proposal, and Section~5 conducts analyses for
both PSQI and EEG data sets. The paper ends with a discussion in
Section~6.



\section{The Low-rank Based Inference Procedure for $\etab$}\label{sec.2}



\subsection{Model specification}

The main idea of our proposal is to utilize the matrix structure of
$\Mb$, from which it is natural to treat $\eta_{jk}$ as the
$(j,k)$-th element of the $p\times q$ matrix $\etab$. With this
matrix representation, model (\ref{model0}) is equivalently
expressed as
\begin{eqnarray}
g\{E(Y|Z,
\Mb)\}=\gamma+\xi^TZ+\vec(\etab)^T\vec(\Mb),\label{model0.m}
\end{eqnarray}
where $\vec(\cdot)$ is the operator that stacks a matrix into a long
vector columnwisely. Under model~(\ref{model0.m}), answering
(Q1)-(Q2) relies on estimating the matrix $\etab$ and testing
\begin{eqnarray}
H_0:\etab=\0_{p\times q}.\label{h0.m}
\end{eqnarray}
To proceed, it is reasonable to assume that most of $\eta_{jk}=0$ in
practice. As a result, $\etab$ is also likely to be a low-rank
matrix, which enables us to impose a parsimonious assumption on
$\etab$ to improve efficiency. It motivates us to consider the
rank-$r$ GLM model
\begin{eqnarray}
g\{E(Y|Z,\Mb)\}=\gamma+\xi^TZ+\vec(\etab)^T\vec(\Mb)\quad{\rm
with}\quad\etab=\Ab\Bb^T,\label{model}
\end{eqnarray}
where $\Ab\in \mathbb{R}^{p\times r}$ and $\Bb\in
\mathbb{R}^{q\times r}$ with a pre-specified $r\le \min\{p,q\}$ such
that rank$(\etab)=r$. One advantage of model~(\ref{model}) is the
parsimony of parameters. In fact, the conventional model
(\ref{model0}) requires $1+m+pq$ parameters, while it is
$1+m+(p+q)r$ for model~(\ref{model}). When $r$ is small, we would
expect an efficiency gain when fitting model~(\ref{model}) and,
hence, a higher detection power to testing (\ref{h0.m}).

Model~(\ref{model}) has been studied in Hung and Wang (2013) (with
the logit link function and $r=1$) and Zhou, Li, and Zhu (2013)
(with arbitrary link function and $r$). Note that using
$\etab=\Ab\Bb^T$ is overparameterized. This can be seen from
$\Ab\Bb^T=\Ab\Cb\Cb^{-1}\Bb^T$ for any nonsingular $\Cb\in
\mathbb{R}^{r\times r}$. As a result, we only require
\begin{eqnarray}
s_r=1+m+(p+q-r)r   \label{no.effect.para}
\end{eqnarray}
parameters for the rank-$r$ model~(\ref{model}). For the sake of
identifiability, both Hung and Wang (2013) and Zhou, Li, and Zhu
(2013) impose extra constraints on $(\Ab,\Bb)$. We should note that
the imposed constraints are not unique. The authors of both papers
also mention that the choice of the constraint relies on the prior
knowledge about the underlying study. Unfortunately, different
constraints will produce different analysis results, which is not a
desired property in practice. To avoid ambiguity, we leave
$(\Ab,\Bb)$ arbitrary to develop our method. This is achievable
since we are interested in $\etab$ instead of $(\Ab,\Bb)$, and only
the identifiability of $\Ab\Bb^T$ is required. We will also show
that the asymptotic properties for the estimator of $\etab$ is
invariant to the choice of the identifiability constraints.
Consequently, we are allowed to use the convenient parameterization
$\etab=\Ab\Bb^T$ without imposing any constraint on $(\Ab,\Bb)$,
which makes our method more applicable in practice.


\begin{rmk}
The idea of treating G$\times$G as a matrix $\Mb=GE^T$ is motivated
from Hung \emph{et al.} (2015), who propose a multistage screening
procedure to identify G$\times$G. Our method is more general in that
the developed inference procedure is also applicable to the case of
binary $Y$. Moreover, we focus on the problem of testing
(\ref{h0.m}), for which the method of Hung \emph{et al.} (2016)
cannot be directly applied.
\end{rmk}

\subsection{Estimation and implementation}

Some notation are defined first for the ease of reference. Let the
data $\{(Y_i,Z_i,\Mb_i)\}_{i=1}^n$ be random copies of $(Y,Z,\Mb)$.
Let $X_i=(1,Z_i^T,\vec(\Mb_i)^T)^T$ be the vector of covariates, and
let $\Xb$ be the $n\times (1+m+pq)$ data matrix with the $i$-th row
being $X_i^T$. Let $\theta=(\gamma,\xi^T,\vec(\Ab)^T,\vec(\Bb)^T)^T$
be the parameters of model~(\ref{model}), and let the induced
parameters of interest be
\begin{eqnarray}
\beta(\theta)=(\gamma,\xi^T, \vec(\Ab\Bb^T)^T)^T,
\end{eqnarray}
which consists of the intercept, the effect of $Z$, and the effect
of $\Mb$.

The log-likelihood function of $\theta$ (apart from constant terms)
is calculated to be
\begin{eqnarray}\label{likelihood.n}
\ell(\theta)=-\frac{1}{2n}\sum_{i=1}^n\{Y_i-\beta(\theta)^TX_i\}^2
\end{eqnarray}
under the normal error model~(\ref{glm.normal}), and is calculated
to be
\begin{eqnarray}\label{likelihood.l}
\ell(\theta)=\frac{1}{n}\sum_{i=1}^nY_i\{\beta(\theta)^TX_i\}-\ln[1+\exp\{\beta(\theta)^TX_i\}].
\end{eqnarray}
under the logistic model~(\ref{glm.binary}). To stabilize the
estimation process, we propose to estimate $\theta$ by the penalized
MLE
\begin{eqnarray}
\widehat\theta=\argmax_{\theta}\Big\{\ell(\theta)-\frac{\lambda}{2}
\|\Ab\|_F^2\|\Bb\|_F^2\Big\},\label{mle_criterion}
\end{eqnarray}
where $\|\cdot\|_F^2$ is the Frobenius norm and $\lambda\geq 0$ is
the penalty. Here we use the penalty $\|\Ab\|_F^2\|\Bb\|_F^2$
instead of the convention $\|\Ab\|_F^2+\|\Bb\|_F^2$. The main reason
is that $\etab=\Ab\Bb^T$ is identifiable while $(\Ab,\Bb)$ are not.
Finally, the parameter of interest is estimated by
\begin{eqnarray}
\widehat \beta=\beta(\widehat\theta)=(\widehat \gamma,\widehat
\xi^T, \vec(\widehat\etab)^T )^T\quad{\rm with}\quad
\widehat\etab=\widehat \Ab \widehat \Bb^T.\label{beta_hat}
\end{eqnarray}
Note that $\widehat\beta$ depends on the values of $(r,\lambda)$.
The rank parameter $r$ determines the accuracy to approximating
$\etab$, and we propose to select $r$ as large as possible while
preserving the estimation efficiency. In particular, we propose to
choose $r$ such that $n/s_r$ is moderately large (e.g., $\ge 5$),
where $s_r$ is the number of effective parameters defined in
(\ref{no.effect.para}). For a fixed $r$, the penalty $\lambda$ is
selected by cross-validation.

To implement our method, we use the alternating method to solve
(\ref{mle_criterion}). First note that model~(\ref{model}) can be
expressed as
\begin{eqnarray}
g\{E(Y|Z)\}&=&\gamma+\xi^TZ+\vec(\Ab)^T\vec(\Mb\Bb)\label{model.B}\\
&=&\gamma+\xi^TZ+\vec(\Bb)^T\vec(\Mb^T\Ab).\label{model.A}
\end{eqnarray}
Observe that (\ref{model.B}) can be treated as the GLM with
parameters $\theta_\Bb=(\gamma,\xi^T,\vec(\Ab)^T)^T$ and data
$\{(Y_i,Z_i,\Mb_i\Bb)\}_{i=1}^n$. Thus, when $\Bb$ is fixed,
maximizing (\ref{mle_criterion}) with respect to $\theta_\Bb$
becomes the conventional penalized MLE problem with the penalty
$\frac{\lambda}{2}\|\Bb\|_F^2$. The case for fixed $\Ab$ is similar
by using the parameters $\theta_\Ab=(\gamma,\xi^T,\vec(\Bb)^T)^T$,
the data $\{(Y_i,Z_i,\Mb_i^T\Ab)\}_{i=1}^n$, and the penalty
$\frac{\lambda}{2}\|\Ab\|_F^2$. We then iterate the roles of $\Ab$
and $\Bb$ until convergence. Detailed implementation algorithm
is summarized below.\\

\hrule  \vspace{0.25cm} \noindent \textbf{Alternating
method}\vspace{0.25cm}\hrule
\begin{enumerate}
\item
Given an initial value $\Bb_{(0)}$ from, e.g., the leading $r$ right
singular vectors of the conventional ridge estimate of $\etab$. For
$k=0,1,2,\ldots,$ do the following Steps~2-3.

\item
Given $\Bb_{(k)}$, obtain the penalized MLE
$(\gamma^*,\xi^*,\Ab_{(k+1)})$ from (\ref{model.B}) using the data
$\{(Y_i,Z_i,\Mb_i\Bb_{(k)})\}_{i=1}^n$ and the penalty
$\frac{\lambda}{2}\|\Bb_{(k)}\|_F^2$.

\item
Given $\Ab_{(k+1)}$, obtain the penalized MLE
$(\gamma_{(k+1)},\xi_{(k+1)},\Bb_{(k+1)})$ from (\ref{model.A})
using the data $\{(Y_i,Z_i,\Mb_i^T\Ab_{(k+1)})\}_{i=1}^n$ and the
penalty $\frac{\lambda}{2}\|\Ab_{(k+1)}\|_F^2$. Define
$\theta_{(k+1)}=(\gamma_{(k+1)},\xi_{(k+1)}^T,\vec(\Ab_{(k+1)}\Bb_{(k+1)}^T)^T)^T$.

\item
Repeat the procedure until the convergence of
$\beta(\theta_{(k+1)})$. Output $\widehat\theta=\theta_{(\infty)}$.
\end{enumerate}
\hrule

~\


\subsection{Asymptotic property}

We now proceed to derive the asymptotic property of $\widehat
\beta$. Let $\beta_0=(\gamma_0,\xi_0^T,\vec(\etab_0)^T)^T$ be the
true value of $\beta$ under model~(\ref{model}), and assume the
existence of a regular point (Shapiro, 1986) $\theta_0$ in the
parameter space of $\theta$ such that $\beta_0=\beta(\theta_0)$.
Define
\begin{eqnarray}
\Deltab(\theta)=\frac{\partial \beta(\theta)}{\partial\theta}=
\left[\begin{array}{ccc}
\Ib_{m+1} & \0 & \0 \\
\0 & (\Bb\otimes \Ib_{p}) & (\Ib_{q}\otimes \Ab)\Kb_{q,r} \\
\end{array}\right],
\end{eqnarray}
where $\Kb_{q,r}$ is the communication matrix satisfying
$\Kb_{q,r}\vec(\Bb)=\vec(\Bb^T)$ for any $q\times r$ matrix $\Bb$,
and let $\Deltab_0=\Deltab(\theta_0)$. The asymptotic property of
$\widehat \beta$ is stated below, where its proof is deferred to
Appendix.


\begin{thm}\label{thm.beta}
Assume the validity of model~(\ref{model}) with
$\beta_0=\beta(\theta_0)$ and ${\rm rank}(\Deltab_0)=s_r$. Assume
also that $\lambda=o_p(n^{-1/2})$. Then, as $n\to\infty$,
$\sqrt{n}(\widehat\beta-\beta_0)\stackrel{d}{\rightarrow}N(0,\Sigmab_0)$
with the asymptotic covariance matrix
$\Sigmab_0=\Deltab_{0}(\Deltab_{0}^{T}\Vb_0\Deltab_{0})^{+}\Deltab_{0}^{T}$,
where $\Vb_0$ is defined below:
\begin{itemize}
\item[(a)]
For normal model~(\ref{glm.normal}), $\Vb_0=\sigma^{-2}
E[X_iX_i^T]$.

\item[(b)]
For logistic model~(\ref{glm.binary}), $\Vb_0=E[\nu_i(\theta_0)
X_iX_i^T]$ with
$\nu_i(\theta)=\frac{\exp\{\beta(\theta)^TX_i\}}{[1+\exp\{\beta(\theta)^TX_i\}]^2}$.
\end{itemize}
\end{thm}

The asymptotic property of $\widehat\beta$ will be the core to
develop our test statistics in Section~\ref{sec.test}, and we
propose to estimate the asymptotic covariance matrix $\Sigmab_0$ by
the sandwich-type estimator
\begin{eqnarray}
\widehat\Sigmab=\widehat\Deltab\left\{\widehat\Deltab^{T}
(\widehat\Vb+\lambda\Ib_{1+m+pq})\widehat\Deltab\right\}^{+}\widehat\Deltab^{T}\widehat\Vb\widehat\Deltab
\left\{\widehat\Deltab^{T}
(\widehat\Vb+\lambda\Ib_{1+m+pq})\widehat\Deltab\right\}^{+}\widehat\Deltab^T,\label{Sigma_normal}
\end{eqnarray}
where $\widehat \Deltab=\Deltab(\widehat\theta)$, $\widehat
\Vb=\widehat\sigma^{-2}\cdot\frac{1}{n}\sum_{i=1}^nX_iX_i^T$ with
$\widehat\sigma^2=\frac{1}{n-s_r}\sum_{i=1}^n(Y_i-\widehat\beta^TX_i)^2$
for the case of (\ref{glm.normal}), and $\widehat
\Vb=\frac{1}{n}\sum_{i=1}^n\nu_i(\widehat\theta)X_iX_i^T$ for the
case of (\ref{glm.binary}). Here we add $\lambda\Ib_{1+m+pq}$ in
(\ref{Sigma_normal}) for the sake of stabilizing estimation, and
note that $\widehat\Sigmab\stackrel{p}{\to}\Sigmab_0$. Consequently,
subsequent inference of $\beta_0$ can be based on
$(\widehat\beta,\widehat\Sigmab)$. For example, an approximated
$100(1-\alpha)$\% confidence interval for the $j$-th element of
$\beta_0$ can be constructed as
\begin{eqnarray}
\widehat\beta_j \pm
\frac{z_{1-\alpha/2}}{\sqrt{n}}[\widehat\Sigmab]^{1/2}_j,
\end{eqnarray}
where $[\widehat\Sigmab]_j$ denotes the $j$-th diagonal element of
$\widehat\Sigmab$, and $z_{1-\alpha/2}$ is the
$(1-\alpha/2)$-quantile of the standard normal distribution.


\section{Detecting the Significance of $\etab$}\label{sec.test}

\subsection{The low-rank based test statistic}

We propose test statistics based on $\widehat\etab$ from the
low-rank model~(\ref{model}) for the null hypothesis (\ref{h0.m}),
which provides a unified estimation-testing procedure for $\etab$.
Motivated from Theorem~\ref{thm.beta}, it is natural to use the
Wald-type test statistic
\begin{eqnarray}
T_{\rm wald}=
\vec(\widehat\etab)^T\left\{[\widehat\Sigmab]_\etab/n\right\}^+\vec(\widehat\etab),\label{test.wald}
\end{eqnarray}
where $[\widehat\Sigmab]_\etab$ is the asymptotic covariance matrix
of $\vec(\widehat\etab)$ from $\widehat\Sigmab$. Note that
$[\widehat\Sigmab]_\etab$ is singular due to over-parameterization
of the low-rank model, and the generalized inverse is used.
Observing that $T_{\rm wald}$ is a weighted sum of the differences
$\vec(\widehat\etab-\0)$, to see if there is enough evidence to
reject $H_0$. This strategy, however, can be less powerful in
testing (\ref{h0.m}) when $\etab$ is sparse, as the contribution of
differences can be averaged out during summation. In view of this
point, an alternative method is to test (\ref{h0.m}) by the test
statistic
\begin{eqnarray}
T_{\rm
max}=\max_{j\in\{m+2,\ldots,1+m+pq\}}\frac{\widehat\beta_j^2}{[\widehat\Sigmab]_j/n},
\label{test.min}
\end{eqnarray}
which is expected to be powerful when $\etab$ is sparse. Note that
$T_{\rm max}$ generalizes the commonly used strategy, the minimum
p-value of $pq$ marginal tests, to test (\ref{h0.m}), in the sense
that $T_{\rm max}$ further considers the joint effects among the
matrix-covariate $\Mb$.

Obviously, $T_{\rm wald}$ and $T_{\rm max}$ have their own merits in
testing (\ref{h0.m}), depending on the sparsity of $\etab$. Ideally,
one should choose the test statistic according to the alternative
hypothesis, which might not be known a priori. A reasonable strategy
is thus to combine $(T_{\rm wald},T_{\rm max})$ to adapt to various
situations. There exists many combination methods based on the
p-values, but they are not applicable in our case since the limiting
distribution of $T_{\rm max}$ is not easy to derive. We thus propose
to combine $(T_{\rm wald},T_{\rm max})$ directly via
\begin{eqnarray}
T=T_{\rm wald}\cdot T_{\rm max}\label{test.combine},
\end{eqnarray}
and a large value of $T$ indicates a rejection of (\ref{h0.m}). The
advantage of the product combination is that $T$ is less affected by
the scale of $T_{\rm wald}$ or $T_{\rm max}$.


To overcome the problem of high-dimensionality in testing
(\ref{h0.m}), GESAT has been proposed with the test statistic
\begin{eqnarray} T_{\rm
gesat}=\Big\|\sum_{i=1}^n(Y_i-\widetilde\gamma-\widetilde\xi^TZ_i)\vec(\Mb_i)\Big\|^2,\label{test.gesat}
\end{eqnarray}
where $(\widetilde\gamma,\widetilde\xi)$ are the restricted MLE of
$(\gamma,\xi)$ under $H_0$. GESAT is shown to be locally most
powerful for $\etab$ in a neighborhood of 0. It thus has superior
performance when $\etab$ has weak effect. However, there is no
guarantee for its performance otherwise.
Considering its
local optimality, it is beneficial to further combine $T$ and
$T_{\rm gesat}$ via
\begin{eqnarray}
T^*=T\cdot T_{\rm gesat}\label{test.combine_g}.
\end{eqnarray}
Our simulation results in Section~\ref{sec.sim} show that, while $T$
and $T_{\rm gesat}$ have comparable performances (depending on the
effect size of $\etab$), $T^*$ is generally the best performer.


\subsection{Calculation of p-values}

Since the null distributions of $T$ and $T^*$ are not
straightforward to derive, we propose to use parametric bootstrap to
obtain their p-values. The main idea of parametric bootstrap is to
generate the null data from model~(\ref{model}) given
$(\gamma,\xi,\etab)=(\widetilde\gamma,\widetilde\xi,\0)$, where
$(\widetilde\gamma,\widetilde\xi)$ are the restricted MLE of
$(\gamma,\xi)$ under $H_0$ (B\r{u}\v{z}kov\'{a}, Lumely, and Rice,
2011). The implementation algorithm is summarized below.



~\

\hrule  \vspace{0.25cm} \noindent \textbf{Parametric bootstrap
test}\vspace{0.2cm}\hrule
\begin{enumerate}
\item
Conditional on $(Z_i,\Mb_i)$, generate $Y_i^{(b)}$ from
model~(\ref{model0.m}) with
$(\gamma,\xi,\etab)=(\widetilde\gamma,\widetilde\xi,\0)$ (and with
$\sigma^2=\frac{1}{n-(m+1)}\sum_{i=1}^n(Y_i-\widetilde\gamma-\widetilde\xi^TZ_i)^2$
for normal error model). Obtain the test statistic $T^{(b)}$
($T^{*(b)}$) by fitting model~(\ref{model}) using
$\{(Y_i^{(b)},Z_i,\Mb_i)\}_{i=1}^n$.

\item
Obtain the p-value as the fraction of $T^{(b)}$'s ($T^{*(b)}$'s)
that exceed $T$ ($T^*$).
\end{enumerate}
\hrule \vspace{0.7cm}

Although parametric bootstrap procedure can be applied in various
situations, its performance depends on the randomness of
$(\widetilde\gamma,\widetilde\xi)$. As a result, using parametric
bootstrap can only control the type-I error asymptotically.
Alternatively, an exact test can be constructed when $Z_i$ vanishes,
i.e., when model~(\ref{model0}) reduces to
\begin{eqnarray}
g\{E(Y|\Mb)\}=\gamma+\vec(\etab)^T\vec(\Mb)\quad{\rm
with}\quad\etab=\Ab\Bb^T.\label{model.no_z}
\end{eqnarray}
In this situation, the null data can be simply generated by randomly
permuting $\Mb_i$ to destroy its connection with $Y_i$. The
implementation algorithm is summarized below.

~\

\hrule  \vspace{0.25cm} \noindent \textbf{Permutation
test}\vspace{0.2cm}\hrule
\begin{enumerate}
\item
Generate $\{\Mb_i^{(b)}\}_{i=1}^n$ by randomly permutating $\{\Mb_i
\}_{i=1}^n$. Obtain the test statistic $T^{(b)}$ ($T^{*(b)}$) by
fitting model~(\ref{model.no_z}) using $\{(Y_i,
\Mb_i^{(b)})\}_{i=1}^n$.

\item
Obtain the p-value as the fraction of $T^{(b)}$'s ($T^{*(b)}$'s)
that exceed $T$ ($T^*$).
\end{enumerate}
\hrule \vspace{0.7cm}

\noindent We remind the readers that the permutation test cannot be
applied in the presence of $Z$. The main reason is that permuting
$\Mb$ destroys not only its connection with $Y$ but also its
connection with $Z$, which further makes the resulting p-value
biased. Both resampling procedures are suggested to obtain the
p-value, according to the underlying data structure.




\section{Simulation Studies}\label{sec.sim}

\subsection{Simulation settings}

Simulation studies are conducted to evaluate our proposal, where we
use the PSQI and EEG data sets (see Section~5 for detailed
descriptions) to generate the simulation data:

\begin{enumerate}
\item[(PSQI)]
Let $Z=(G^T,E^T)^T$ and $\Mb=GE^T$, where $G\in \mathbb{R}^{15}$ and
$E\in \mathbb{R}^7$ are randomly generated from the PSQI data (with
sample size $n=400$). Given $(Z,\Mb)$, $Y$ is generated from normal
error model~(\ref{model0.m}) with $\gamma=10$, $\xi=(\xi_{G}^T,
\0_{p-5}^T, \xi_{E}^T, \0_{q-3}^T)^T$, $\xi_{G}=\1_{5}$, and
$\xi_{E}=\1_{3}$, and the specification of $\etab$ is described
separately. In this case, the normal error model~(\ref{model}) with
$r=3$ is fitted.

\item[(EEG)]
The matrix $\Mb\in \mathbb{R}^{6\times 6}$ is randomly generated
from the EEG data (with sample size $n=150$). Given $\Mb$, $Y$ is
generated from logistic model~(\ref{model0.m}) and
(\ref{model.no_z}) with $\gamma=0$, and the specification of $\etab$
is described separately. In this case, the logistic
model~(\ref{model}) with $r=2$ is fitted.
\end{enumerate}
Simulation results are reported with 500 replicates.

\subsection{Simulation results: part-1}

We evaluate the performances of $(\widehat\beta,\widehat\Sigmab)$
under
\begin{eqnarray*}
\etab =\left[\begin{array}{cc}
\etab_{1}  &   \textbf{0}_{2 \times{ (q-1)} }\\
 \textbf{0}_{(p-2) \times 1}  &    \textbf{0}_{(p-2) \times (q-1)}  \\
\end{array}\right]\quad{\rm with}\quad
\etab_{1}=\frac{1}{\sqrt{2}}\cdot\1_2.
\end{eqnarray*}
Simulation results are reported in Table~\ref{table.ge}, which
provides the means and standard deviations (SD) of $\widehat\beta$,
and standard errors (SE) from the means of the diagonal elements of
$\widehat\Sigmab$ that correspond to $(\gamma,\xi_G,\xi_E,\etab_1)$,
and report the averaged mean square error (AMSE) of
$(\widehat\beta_{j}-0)^2$ over the rest parameters (with zero
values) to summarize the performance of $\widehat\beta$. One can see
that the biases of $\widehat\beta$ arise under both PSQI and EEG
settings, but they are relatively small in comparison with the
corresponding SDs. Together with a small value of AMSE,
$\widehat\beta$ is demonstrated to be a consistent estimator of
$\beta_0$. Moreover, a similar values of SDs and SEs support the
validity of $\widehat\Sigmab$ in estimating $\Sigmab_0$. It implies
that the asymptotic property of the low-rank model~(\ref{model})
approximately holds even when the sample size is not large. In fact,
it only requires 80 parameters for model~(\ref{model}) with $r=3$
instead of 128 in the conventional case, which makes the large
sample theory more plausible to be valid.



\subsection{Simulation results: part-2}

\begin{figure}[!ht]
\centering
\includegraphics[width=3in]{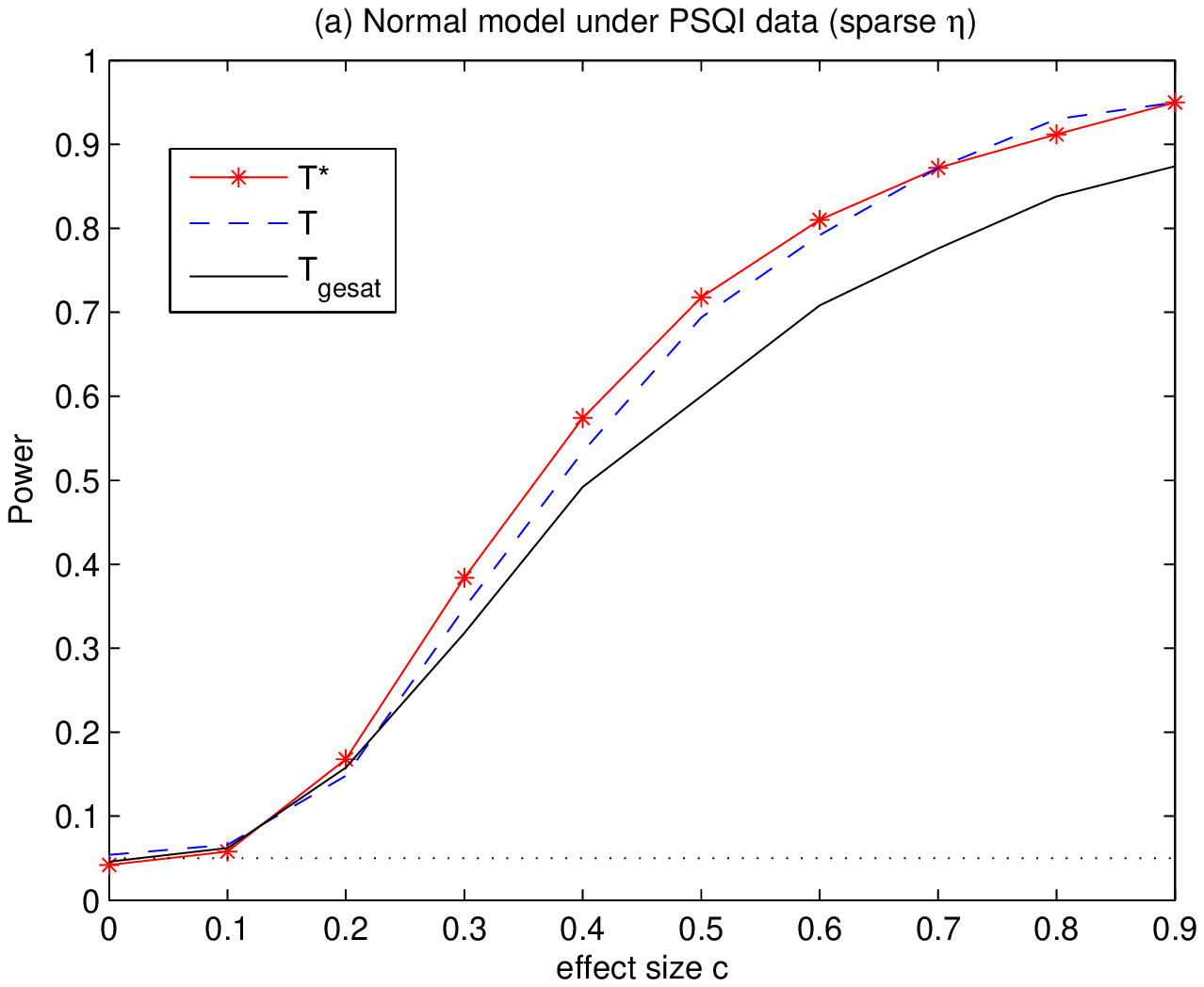}
\includegraphics[width=3in]{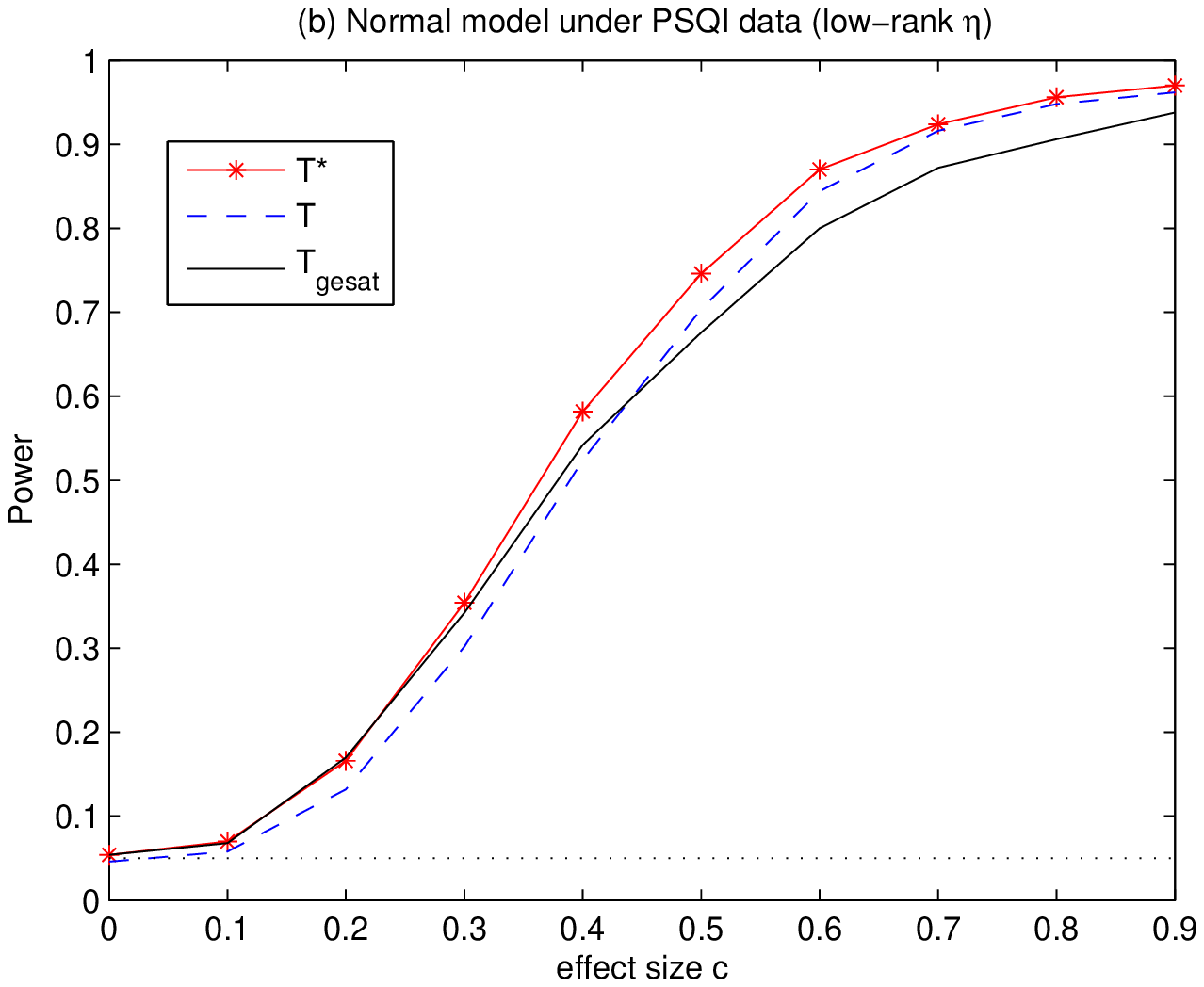}
\includegraphics[width=3in]{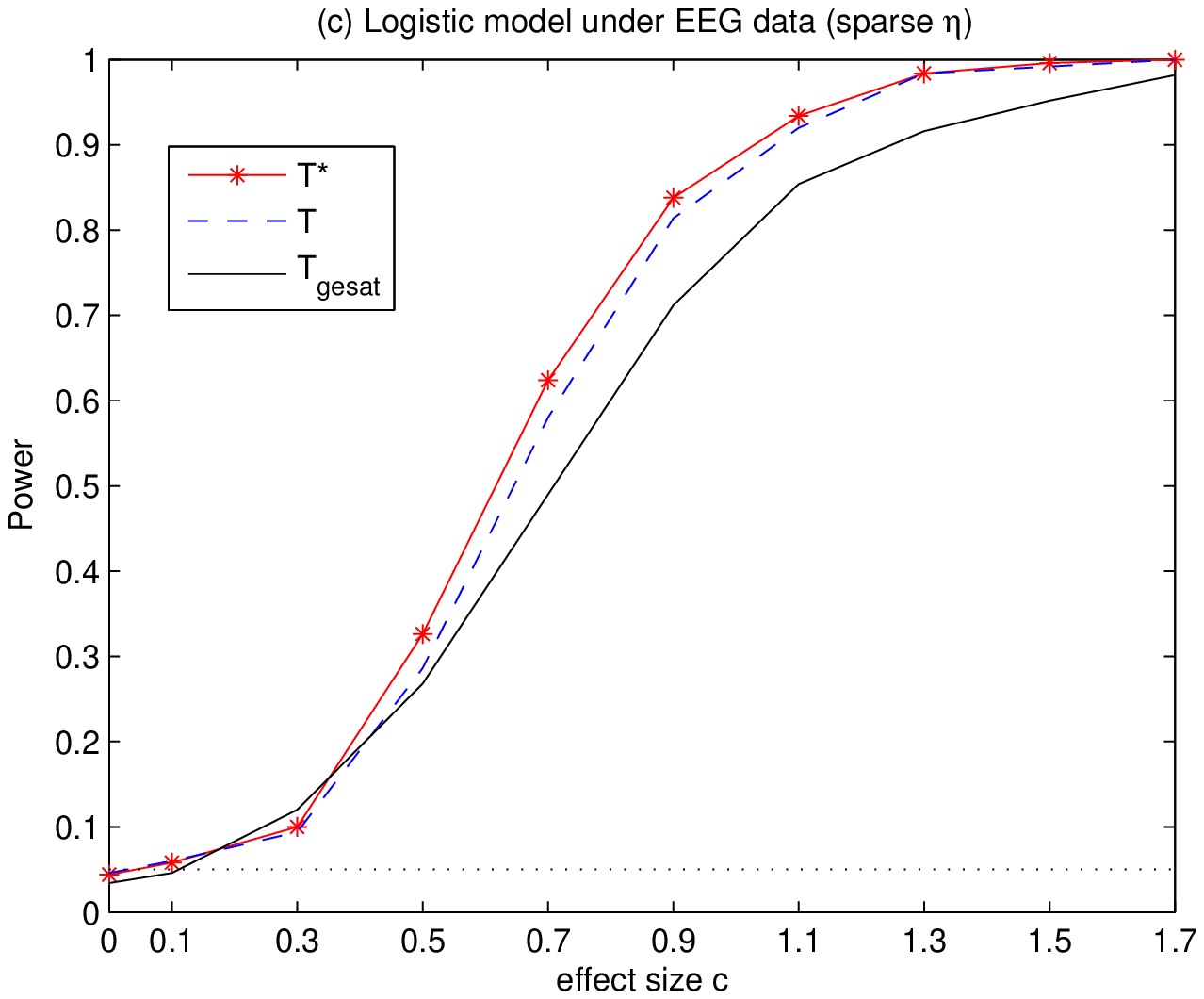}
\includegraphics[width=3in]{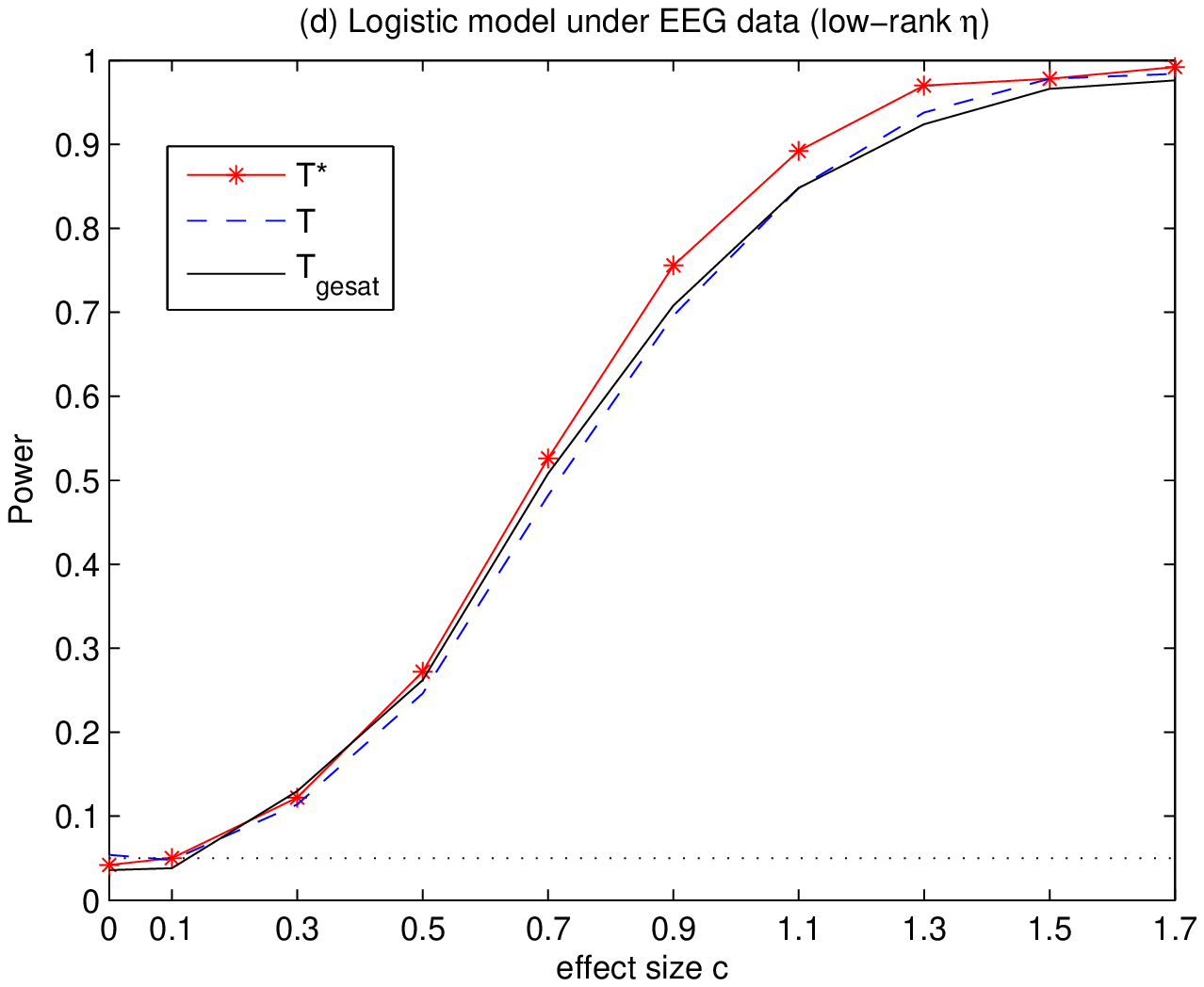}
\caption{The power functions under two types of $\etab$ (sparse in
the left panel, and low-rank in the right panel) at different effect
sizes $c$. (a)-(b): the normal model using the PSQI data; (c)-(d):
the logistic model using the EEG data. The dotted horizontal lines
represents the significance level 0.05.}\label{fig.power}
\end{figure}

This simulation evaluates the detection powers of $T$ and $T^*$ (at
the significance level 0.05) under two types of $\etab$:
\begin{itemize}
\item
\textbf{Sparse $\etab$}: In each simulation run, $\etab$ has zero
effects except for $2$ randomly selected elements. The values of the
$2$ selected elements are generated from $c\cdot U$, where
$U\in\mathbb{R}^2$ is randomly generated from unit sphere and $c$ is
the effect size.

\item
\textbf{Low-rank $\etab$}: $\etab$ has zero effects except for the
first two columns. For each simulation run, the non-zero elements
are generated from $c\cdot U$, where $U\in\mathbb{R}^{2p}$ is
randomly generated from unit sphere and $c$ is the effect size.
\end{itemize}
The penalty $\lambda$ is selected by cross-validation over
$\{\frac{s_r}{n^{3/2}},\frac{s_r}{n},\frac{s_r}{\sqrt{n}\log(n)}\}$
(such that the condition $\lambda=o_p(n^{-1/2})$ is satisfied),
where $s_r$ is defined in (\ref{no.effect.para}).
Figure~\ref{fig.power} reports the power functions of $T$, $T^*$,
and $T_{\rm gesat}$ for testing (\ref{h0.m}) at different effect
sizes $c$. For fair comparisons, we use the same re-sampling scheme
to obtain the p-value of $T_{\rm gesat}$. One can see that all
methods correctly control the type-I errors at 0.05 level, which
implies the validity of the re-sampling schemes (parametric
bootstrap or permutation) to obtain valid p-values.

Comparing the detection powers for small effect size $c$, $T$ and
$T_{\rm gesat}$ are detected to have comparable performances when
$\etab$ is sparse, and $T$ is slightly worse than $T_{\rm gesat}$
when $\etab$ is low-rank, while $T$ outperforms $T_{\rm gesat}$ when
$c$ is moderate to large. This is reasonable since the performance
of $T_{\rm gesat}$ can only be guaranteed for $\etab$ in a
neighborhood of $\0$. We also observe that $T$ has a significant
improvement over $T_{\rm gesat}$ when $\etab$ is sparse, which
indicates the usefulness of incorporating (\ref{test.min}) in $T$.
For the case of low-rank $\etab$, the improvement of $T$ is observed
for normal error model, while the difference between $T$ and $T_{\rm
gesat}$ vanishes for logistic model. Although $T$ and $T_{\rm
gesat}$ have comparable performances, we should note that $T$ is
directly associated with $\widehat\etab$, from which a unified
estimation-testing procedure for $\etab$ is available, while this is
not the case for $T_{\rm gesat}$.

The above observations reflect the fact that both $T$ and $T_{\rm
gesat}$ have their own merits in detecting (\ref{h0.m}), depending
on the underlying structure of $\etab$. Moreover, by combing both
methods, $T^*$ is detected to be the best performer in all
situations. In summary, our simulations indicate the applicabilities
of $T^*$ regardless of the forms of $\etab$. It also indicates that,
by imposing a low-rank assumption, we can further improve existing
methods (such as GESAT) that ignore the matrix structure of $\etab$.

\section{Data Analyses}

\subsection{The PSQI data}\label{sec.psqi}

The PSQI data set (Lai \emph{et al}., 2014) includes 359 subjects
with an average age of 41 years old (ranges from 18 to 69). The
participants consist of 214 females and 145 males. For each subject,
markers on 3 genes are collected: RORA (2 markers), RORB(17
markers), and NR1D1 (4 markers). Also collected for each subject are
the assessments of sleep quantity from the Pittsburgh Sleep Quality
Index (PSQI) of Buysse \emph{et al}. (1989), which consists of seven
scores: sleep quality, sleep latency, sleep duration, habitual sleep
efficiency, sleep disturbance, use of sleeping medication, and
daytime dysfunction.

In our analysis, we consider the response $Y$ to be the
log-transformation of the sum of all PSQI scores (plus one to avoid
taking logarithm for 0), and let ROR contains RORA and RORB, which
has 19 markers. Let $\Mb$ be the $19\times 4$ matrix of the
interactions ROR$\times$NR1D1, and let $Z$ consist of age and gender
as possible confounding factors. Our interest focuses on whether
ROR$\times$NR1D1 are associated with the PSQI scores. Fitting the
normal error model~(\ref{model}) with $r=3$ gives the p-values (from
parametric bootstrap test) of $T$, $T^*$, and $T_{\rm gesat}$ to be
0.001, 0.006, and 0.173, respectively. Thus, only the low-rank based
methods declare that ROR$\times$NR1D1 are influential to PSQI score.
A large p-value of $T_{\rm gesat}$ also indicates that the true
$\etab$ may have moderate effect size, so that $T_{\rm gesat}$ has
limited detection power to declare its significance.

To further identify the root causes, we report the estimated effect
sizes of ROR$\times$NR1D1 in Table~\ref{pvaluetable.rabnr}, where
the significant effects (identified by the parametric bootstrap test
p-values of $\widehat\beta_j$'s after Bonferroni correction) are
marked as bold. In particular, 3 specific interactions among
rs1144047, rs1327836, rs2269457, and rs12941497 are identified,
which can be the candidate interactions for further investigations.
It also indicates a sparse effects of ROR$\times$NR1D1 on the PSQI
score.

\subsection{The EEG data}\label{sec.eeg}

The EEG data contains 77 subjects in alcoholic group ($Y=1$) and 45
subjects in control group ($Y=0$). The measurement of voltage values
at $256$ time points and $64$ channels are collected for each
subject. It is interesting in investigating if the EEG signal is
associated with the alcoholic status $Y$. The raw data can be
obtained from the UCI Machine Learning Repository
(\textsf{http://archive.ics.uci.edu/ml/datasets/EEG+Database}).

In our analysis, we use the same data as the one used in Hung and
Wang (2013), and follow a similar procedure for data pre-processing.
In particular, we first reduce the dimensionality of the original
$256\times64$ matrix-covariates to $10\times 10$ by multilinear
principal component analysis (MPCA). See Hung \emph{et al.} (2012)
for an illustration of MPCA. The matrix $\Mb$ is then obtained from
the reduced $10\times10$ matrix after componentwise standardization.
Based on $(Y,\Mb)$, the logistic model~(\ref{model}) with $r=2$ is
fitted. In this data analysis, $T$, $T^*$, and $T_{\rm gesat}$ all
produce  p-values (from permutation test) $<10^{-3}$, indicating a
strong association between the EEG signals and the alcoholic status.
This result confirms the findings from Hung and Wang (2013) and
Zhou, Li, and Zhu (2013), who both analyze the EEG data.

To delve into the association between $\Mb$ and $Y$, we report the
estimates $\widehat\etab$ in Table~\ref{pvaluetable.eegeffect},
where the significant elements (identified by the permutation test
p-values of $\widehat\beta_j$'s after Bonferroni correction) are
marked as bold. One can see that most of the effective elements of
$\Mb$ are in the position of its second row. It indicates the
influence of the EEG signals should have a low-rank structure, where
only a few channels can have functions to alcoholic status over
times. We remind the readers that our estimates $\widehat\etab$ is
obtained without imposing any identifiability constraint for the
parameterization $\etab=\Ab\Bb^T$, while the results from Hung and
Wang (2013) and Zhou, Li, and Zhu (2013) do.

\section{Conclusion}

In this paper, we propose powerful methods to detect the
significance of matrix-covariate. We utilize the matrix structure of
$\etab$ to achieve a parsimonious parameterization and, hence, a
higher detection power than conventional methods. Our proposal is
different from existing methods in that no identifiability
constraint is required when making inference about $\etab$. Another
advantage is that our method can provide estimates of $\etab$ at the
same time. By reporting $\widehat\etab$, we identify a sparse effect
of G$\times$G in the PSQI data, while a low-rank effect is detected
in the EEG data.

The matrix-covariate discussed in this work is an order-two tensor.
Tensor-covariate can be found in many applications nowadays. For
example, $p$ covariates measured at $q$ time points under $k$
environments corresponds to an order-three tensor (with dimension
$p\times q\times k$) for each subject. Statistical inference
procedure for GLM with tensor-covariate has been developed in Zhou,
Li, and Zhu (2013) and Zhou and Li (2014), where the authors focus
on the estimation of the effect size of tensor-covariate. When the
research aim is to test the existence of association between the
response and tensor-covariate, our low-rank based test statistics
$T$ and $T^*$ can be extended, provided that a version of the
asymptotic property Theorem~\ref{thm.beta} is developed for
tensor-covariate. Another issue is the chosen ``low-rank''
parameterization of a tensor, which is not unique as in the case of
matrix. This can be a future study.


~\ \\

\begin{center}
{\large REFERENCES}
\end{center}
\begin{description}
\item
Buysse, D. J., Reynolds, C. F., and Monk, T. H. (1989). The
Pittsburgh sleep quality index: A new instrument for psychiatric
practice and research. \emph{Psychiatry Research} 28, 193-213.

\item
B\r{u}\v{z}kov\'{a}, P., Lumely, T., and Rice., K. (2011).
Permutation and parametric bootstrap tests for gene-gene and
gene-environment interactions. \emph{Annals of Human Genetics} 75,
36-45.

\item
Lin, X. (1997). Variance component testing in generalised linear
models with random effects. \emph{Biometrika} 84, 309-326.

\item
Lin, X., Lee, S., Christiani, D. C., and Lin, X. (2013). Test for
interactions between a genetic marker set and environment in
generalized linear models. \emph{Biostatistics} 14, 667-681.

\item
Henderson, H. V. and Searle, S. R. (1979). Vec and vech operators
for matrices, with some uses in Jacobians and multivariate
statistics. Canadian Journal of Statistics, 7, 65-81.


\item
Hung, H., Wu, P. S., Tu, I. P., and Huang, S. Y. (2012). On
multilinear principal component analysis of order-two tensors.
\textit{Biometrika} 99, 569-583.

\item
Hung, H. and Wang, C.~C. (2013). Matrix variate logistic regression
model with application to EEG data. \emph{Biostatistics} 14,
189-202.

\item
Hung, H., Lin, Y. T., Wang, C. C., Chen, P., Huang, S. Y., and
Tzeng, J. Y. (2015). Detection of gene-gene interactions using
multistage sparse and low-rank regression. \emph{Biometrics}. (doi:
10.1111/biom.12374)

\item
Lai, Y. C., Huang, M. C., Chen, H. C., Lu, M. K., Chiu, Y. H., Shen,
W. W., Lu, R. B., and Kuo, P. H. (2014). Familiality and clinical
outcomes of sleep disturbance in major depressive and bipolar
disorders. \textit{Journal of Psychosomatic Research} 76, 61-67.

\item
Shapiro, A. (1986). Asymptotic theory of overparameterized
structural models. \emph{Journal of American Statistical
Association}, 81, 142-149.

\item
Zhou, H., Li, L., and Zhu, H. (2013). Tensor regression with
applications in neuroimaging data analysis. \emph{Journal of the
American Statistical Association} 108, 540-552.

\item
Zhou, H. and Li, L. (2014). Regularized matrix regression.
\emph{Journal of the Royal Statistical Society: Series B}, 76,
463-483.

\end{description}

\newpage

\newpage

\begin{table}
\centering \caption{The means (Mean) and standard deviations (SD) of
$\hat{\beta}$, and the standard errors (SE) from the diagonal
elements of $\widehat\Sigmab$. The last row gives the mean and
standard deviation of AMSE.}
 \vspace{3ex}
\begin{tabular}{lrrrrrrrrrrrrrrrrrrr}
\hline
&& \multicolumn{2}{c}{PSQI} && \multicolumn{5}{c}{EEG}\\
\cline{2-5} \cline{7-10} 

&\multicolumn{1}{c}{True}  &\multicolumn{1}{c}{Mean} &\multicolumn{1}{c}{SD} &\multicolumn{1}{c}{SE} &&\multicolumn{1}{c}{True}  &\multicolumn{1}{c}{Mean} &\multicolumn{1}{c}{SD} &\multicolumn{1}{c}{SE} \\
\hline
        $\gamma$    &   10.000  &   9.998  &   0.062   &   0.059        &&   0.000   &   -0.002  &   0.212   &   0.205   \\
        $\xi_{G}$   &   1.000   &   0.996   &   0.107   &   0.104    \\
                    &   1.000   &   0.986   &   0.113   &   0.104    \\
                    &   1.000   &   1.014   &   0.108   &   0.103    \\
                    &   1.000   &   1.005   &   0.112   &   0.103    \\
                    &   1.000   &   0.995   &   0.111   &   0.102    \\
        $\xi_{E}$   &   1.000   &   1.005   &   0.085   &   0.081    \\
                    &   1.000   &   0.994   &   0.091   &   0.081    \\
                    &   1.000   &   0.998   &   0.082   &   0.080    \\
        $\etab_{1}$ &   0.707   &   0.638   &   0.117   &   0.108    &&   0.707   &   0.657   &   0.251   &   0.220   \\
                    &   0.707   &   0.631   &   0.126   &   0.108    &&   0.707   &   0.634   &   0.284   &   0.233   \\
\hline
    AMSE    &      &   0.008   &   0.002                            &&&      &   0.036 & 0.013 \\
\hline
\end{tabular}
\vspace{5ex} \label{table.ge}
\end{table}

\clearpage

\begin{table}[h]
\centering \caption{ The effect sizes $\widehat\etab$ of
ROR$\times$NR1D1 from normal model~(\ref{model}) with $r=3$, where
the significant elements are marked as bold. }
 \vspace{3ex}
\begin{tabular}{llcccccccccccccccccccccc}

\hline

  &\multicolumn{5}{c}{NR1D1}\\
\cline{2-6}
        &   &   rs2314339   &   rs2071427   &   rs2269457   &   rs12941497          \\
    \hline

    RORA    &   rs809736    &   -0.007  &   -0.026  &   -0.019  &   0.047   \\
        &   rs4774388   &   0.001   &   0.014   &   0.016   &   -0.034  \\
    RORB    &   rs10491929  &   0.002   &   0.004   &   0.002   &   -0.005  \\
        &   rs17611535  &   0.001   &   0.015   &   0.017   &   -0.036  \\
        &   rs10217594  &   -0.004  &   0.007   &   0.016   &   -0.030  \\
        &   rs7037043   &   -0.008  &   -0.023  &   -0.014  &   0.036   \\
        &   rs2025882   &   -0.001  &   -0.004  &   -0.003  &   0.006   \\
        &   rs7022435   &   -0.001  &   -0.005  &   -0.004  &   0.010   \\
        &   rs3750420   &   0.001   &   0.001   &   0.000   &   -0.001  \\
        &   rs1013078   &   -0.004  &   -0.027  &   -0.026  &   0.059   \\
        &   rs2273975   &   -0.004  &   0.012   &   0.023   &   -0.044  \\
        &   rs3903529   &   0.003   &   -0.004  &   -0.011  &   0.020   \\
        &   rs11144041  &   0.004   &   0.006   &   0.000   &   -0.003  \\
        &   rs7021908   &   -0.002  &   0.000   &   0.004   &   -0.006  \\
        &   rs7865407   &   -0.003  &   -0.012  &   -0.009  &   0.022   \\
        &   rs11144047  &   -0.007  &   -0.040  &   -0.036  &   $\mathbf{0.083}$    \\
        &   rs1327836   &   0.005   &   0.054   &   $\mathbf{0.058}$    &   -$\mathbf{0.125}$   \\
        &   rs11144064  &   -0.001  &   -0.005  &   -0.004  &   0.010   \\
        &   rs4098048   &   0.003   &   0.000   &   -0.005  &   0.008   \\
\hline
\end{tabular}
\vspace{5ex} \label{pvaluetable.rabnr}
\end{table}

\begin{table}[h]
\vspace{3ex} \centering \caption{The effect sizes $\widehat\etab$ of
EEG from logistic model~(\ref{model}) with $r=2$, where the
significant effects are marked as bold.}
 \vspace{3ex}
\begin{tabular}{ccccccccccccccccccccccc}

\hline
                0.063   &   -0.113  &   0.059   &   -0.026  &   -0.008  &   -0.015  &   -0.029  &   -0.009  &   0.031   &   -0.035  \\
                -$\mathbf{0.108}$   &   $\mathbf{0.217}$    &   -$\mathbf{0.133}$   &   0.068   &   -0.023  &   0.018   &   0.016   &   -0.070  &   -0.073  &   0.112   \\
                0.074   &   -$\mathbf{0.170}$   &   $\mathbf{0.119}$    &   -0.067  &   0.047   &   -0.005  &   0.018   &   0.120   &   0.068   &   -$\mathbf{0.121}$   \\
                0.004   &   -0.021  &   0.022   &   -0.015  &   0.020   &   0.003   &   0.017   &   0.046   &   0.014   &   -0.031  \\
                0.049   &   -0.102  &   0.064   &   -0.034  &   0.015   &   -0.007  &   -0.003  &   0.042   &   0.036   &   -0.057  \\
                0.056   &   -0.110  &   0.066   &   -0.033  &   0.008   &   -0.010  &   -0.012  &   0.028   &   0.036   &   -0.053  \\
                0.008   &   -0.042  &   0.045   &   -0.031  &   0.041   &   0.007   &   0.035   &   $\mathbf{0.096}$    &   0.028   &   -0.064  \\
                -0.033  &   0.068   &   -0.042  &   0.022   &   -0.008  &   0.005   &   0.004   &   -0.024  &   -0.023  &   0.036   \\
                -0.021  &   0.050   &   -0.035  &   0.020   &   -0.014  &   0.002   &   -0.006  &   -0.036  &   -0.020  &   0.036   \\
                0.038   &   -0.067  &   0.034   &   -0.015  &   -0.006  &   -0.009  &   -0.019  &   -0.008  &   0.018   &   -0.019  \\

\hline

\end{tabular}

\vspace{5ex} \label{pvaluetable.eegeffect}
\end{table}

\end{document}